\PassOptionsToPackage{unicode}{hyperref}
\PassOptionsToPackage{hyphens}{url}
\PassOptionsToPackage{dvipsnames,svgnames,x11names}{xcolor}
\documentclass[
]{article}
\usepackage{amsmath,amssymb}
\usepackage{lmodern}
\usepackage{iftex}
\ifPDFTeX
  \usepackage[T1]{fontenc}
  \usepackage[utf8]{inputenc}
  \usepackage{textcomp} 
\else 
  \usepackage{unicode-math}
  \defaultfontfeatures{Scale=MatchLowercase}
  \defaultfontfeatures[\rmfamily]{Ligatures=TeX,Scale=1}
\fi
\IfFileExists{upquote.sty}{\usepackage{upquote}}{}
\IfFileExists{microtype.sty}{
  \usepackage[]{microtype}
  \UseMicrotypeSet[protrusion]{basicmath} 
}{}
\makeatletter
\@ifundefined{KOMAClassName}{
  \IfFileExists{parskip.sty}{%
    \usepackage{parskip}
  }{
    \setlength{\parindent}{0pt}
    \setlength{\parskip}{6pt plus 2pt minus 1pt}}
}{
  \KOMAoptions{parskip=half}}
\makeatother
\usepackage{xcolor}
\usepackage{color}
\usepackage{fancyvrb}

\DefineVerbatimEnvironment{Highlighting}{Verbatim}{commandchars=\\\{\}}
\newenvironment{Shaded}{}{}

\newcommand{\BuiltInTok}[1]{\textcolor[rgb]{0.00,0.50,0.00}{#1}}

\newcommand{\CommentTok}[1]{\textcolor[rgb]{0.38,0.63,0.69}{\textit{#1}}}

\newcommand{\DecValTok}[1]{\textcolor[rgb]{0.25,0.63,0.44}{#1}}

\newcommand{\FloatTok}[1]{\textcolor[rgb]{0.25,0.63,0.44}{#1}}

\newcommand{\ImportTok}[1]{\textcolor[rgb]{0.00,0.50,0.00}{\textbf{#1}}}

\newcommand{\NormalTok}[1]{#1}
\newcommand{\OperatorTok}[1]{\textcolor[rgb]{0.40,0.40,0.40}{#1}}

\newcommand{\StringTok}[1]{\textcolor[rgb]{0.25,0.44,0.63}{#1}}

\newcommand{\VerbatimStringTok}[1]{\textcolor[rgb]{0.25,0.44,0.63}{#1}}

\usepackage{graphicx}
\makeatletter
\def\maxwidth{\ifdim\Gin@nat@width>\linewidth\linewidth\else\Gin@nat@width\fi}
\def\maxheight{\ifdim\Gin@nat@height>\textheight\textheight\else\Gin@nat@height\fi}
\makeatother
\setkeys{Gin}{width=\maxwidth,height=\maxheight,keepaspectratio}
\makeatletter
\def\fps@figure{htbp}
\makeatother
\NewDocumentCommand\citeproctext{}{}
\NewDocumentCommand\citeproc{mm}{%
  \begingroup\def\citeproctext{#2}\cite{#1}\endgroup}
\makeatletter
 \let\@cite@ofmt\@firstofone
 \def\@biblabel#1{}
 \def\@cite#1#2{{#1\if@tempswa , #2\fi}}
\makeatother
\newlength{\cslhangindent}
\newlength{\csllabelwidth}
\newenvironment{CSLReferences}[2] 
 {\begin{list}{}{%
  \setlength{\itemindent}{0pt}
  \setlength{\leftmargin}{0pt}
  \setlength{\parsep}{0pt}
  \ifodd #1
   \setlength{\leftmargin}{\cslhangindent}
   \setlength{\itemindent}{-1\cslhangindent}
  \fi
  \setlength{\itemsep}{#2\baselineskip}}}
 {\end{list}}
\usepackage{calc}

\ifLuaTeX
\usepackage[bidi=basic]{babel}
\else
\usepackage[bidi=default]{babel}
\fi
\babelprovide[main,import]{american}

\def\languageshorthands#1{}
\ifLuaTeX
  \usepackage{selnolig}  
\fi
\IfFileExists{bookmark.sty}{\usepackage{bookmark}}{\usepackage{hyperref}}
\IfFileExists{xurl.sty}{\usepackage{xurl}}{} 
\hypersetup{
  pdftitle={bayes\_spec: A Bayesian Spectral Line Modeling Framework for
Astrophysics},
  pdfauthor={Trey V. Wenger},
  pdflang={en-US},
  colorlinks=true,
  linkcolor={Maroon},
  filecolor={Maroon},
  citecolor={Blue},
  urlcolor={Blue},
  pdfcreator={LaTeX via pandoc}}

\title{bayes\_spec: A Bayesian Spectral Line Modeling Framework for
Astrophysics}

\definecolor{c53baa1}{RGB}{83,186,161}
\definecolor{c202826}{RGB}{32,40,38}

\usepackage[affil-it]{authblk}
\usepackage{orcidlink}
\author[1%
  ]{Trey V. Wenger%
    \,\orcidlink{0000-0003-0640-7787}\,%
    }

\affil[1]{NSF Astronomy \& Astrophysics Postdoctoral Fellow, University
of Wisconsin-Madison, USA%
  }
\date{15 August 2024}

\begin{document}
\maketitle

\section{Summary}\label{summary}

The study of the interstellar medium (ISM)---the matter between the
stars---relies heavily on the tools of spectroscopy. Spectral line
observations of atoms, ions, and molecules in the ISM reveal the
physical conditions and kinematics of the emitting gas. Robust and
efficient numerical techniques are thus necessary for inferring the
physical conditions of the ISM from observed spectral line data.

\texttt{bayes\_spec} is a Bayesian spectral line modeling framework for
astrophysics. Given a user-defined model and a spectral line dataset,
\texttt{bayes\_spec} enables inference of the model parameters through
different numerical techniques, such as Monte Carlo Markov Chain (MCMC)
methods, implemented in the PyMC probabilistic programming library
(\citeproc{ref-pymc2023}{Oriol et al., 2023}). The API for
\texttt{bayes\_spec} is designed to support astrophysical researchers
who wish to ``fit'' arbitrary, user-defined models, such as simple
spectral line profile models or complicated physical models that include
a full physical treatment of radiative transfer. These models are
``cloud-based'', meaning that the spectral line data are decomposed into
a series of discrete clouds with parameters defined by the user's model.
Importantly, \texttt{bayes\_spec} provides algorithms to determine the
optimal number of clouds for a given model and dataset.

\section{Statement of need}\label{statement-of-need}

Bayesian models of spectral line observations are rare in astrophysics.
Physical inference is typically achieved through inverse modeling: the
spectral line data are decomposed into Gaussian components, and then the
physical parameters are inferred from the fitted Gaussian parameters
under numerous assumptions. For example, this is the strategy of
\texttt{gausspy} (\citeproc{ref-lindner2015}{Lindner et al., 2015}),
\texttt{ROHSA} (\citeproc{ref-marchal2019}{Marchal et al., 2019}),
\texttt{pyspeckit} (\citeproc{ref-ginsburg2022}{Ginsburg et al., 2022}),
and \texttt{MWYDYN} (\citeproc{ref-rigby2024}{Rigby et al., 2024}). This
strategy suffers from two problems: (1) the degeneracies of Gaussian
decomposition and (2) the assumptions of the inverse model. Bayesian
forward models, like those enabled by \texttt{bayes\_spec}, can overcome
both of these limitations because (1) prior knowledge about the physical
conditions can reduce the space of possible solutions, and (2) all
assumptions are explicitly built into the model rather than being
applied \emph{a priori}.

\texttt{bayes\_spec} is inspired by
\href{https://github.com/AnitaPetzler/AMOEBA}{AMOEBA}
(\citeproc{ref-petzler2021}{Petzler et al., 2021}), an MCMC-based
Bayesian model for interstellar hydroxide observations. \texttt{McFine}
(\citeproc{ref-williams2024}{Williams \& Watkins, 2024}) is a new
MCMC-based Bayesian model for hyperfine spectroscopy similar in spirit
to \texttt{bayes\_spec}. With \texttt{bayes\_spec}, we aim to provide a
user-friendly, general-purpose Bayesian modeling framework for
\emph{any} astrophysical spectral line observation.

\section{Usage}\label{usage}

Here we demonstrate how to use \texttt{bayes\_spec} to fit a simple
Gaussian line profile model to a synthetic spectrum. For more details,
see the \href{https://bayes-spec.readthedocs.io}{documentation and
tutorials}.

\begin{Shaded}
\begin{Highlighting}[]
\CommentTok{\# Generate data structure}
\ImportTok{import}\NormalTok{ numpy }\ImportTok{as}\NormalTok{ np}
\ImportTok{from}\NormalTok{ bayes\_spec }\ImportTok{import}\NormalTok{ SpecData}

\NormalTok{velocity\_axis }\OperatorTok{=}\NormalTok{ np.linspace(}\OperatorTok{{-}}\FloatTok{250.0}\NormalTok{, }\FloatTok{250.0}\NormalTok{, }\DecValTok{501}\NormalTok{) }\CommentTok{\# km s{-}1}
\NormalTok{noise }\OperatorTok{=} \FloatTok{1.0} \CommentTok{\# K}
\NormalTok{brightness\_data }\OperatorTok{=}\NormalTok{ noise }\OperatorTok{*}\NormalTok{ np.random.randn(}\BuiltInTok{len}\NormalTok{(velocity\_axis)) }\CommentTok{\# K}
\NormalTok{observation }\OperatorTok{=}\NormalTok{ SpecData(}
\NormalTok{    velocity\_axis, brightness\_data, noise,}
\NormalTok{    ylabel}\OperatorTok{=}\VerbatimStringTok{r"Brightness Temperature $T\_B$ (K)"}\NormalTok{,}
\NormalTok{    xlabel}\OperatorTok{=}\VerbatimStringTok{r"LSR Velocity $V\_\{\textbackslash{}rm LSR\}$ (km s$\^{}\{{-}1\})$"}\NormalTok{,}
\NormalTok{)}
\NormalTok{data }\OperatorTok{=}\NormalTok{ \{}\StringTok{"observation"}\NormalTok{: observation\}}

\CommentTok{\# Prepare a three cloud GaussLine model with polynomial baseline degree = 2}
\ImportTok{from}\NormalTok{ bayes\_spec.models }\ImportTok{import}\NormalTok{ GaussModel}

\NormalTok{model }\OperatorTok{=}\NormalTok{ GaussModel(data, n\_clouds}\OperatorTok{=}\DecValTok{3}\NormalTok{, baseline\_degree}\OperatorTok{=}\DecValTok{2}\NormalTok{)}
\NormalTok{model.add\_priors()}
\NormalTok{model.add\_likelihood()}

\CommentTok{\# Evaluate the model for a given set of parameters to generate a}
\CommentTok{\# synthetic "observation"}
\NormalTok{sim\_brightness }\OperatorTok{=}\NormalTok{ model.model.observation.}\BuiltInTok{eval}\NormalTok{(\{}
    \StringTok{"fwhm"}\NormalTok{: [}\FloatTok{25.0}\NormalTok{, }\FloatTok{40.0}\NormalTok{, }\FloatTok{35.0}\NormalTok{], }\CommentTok{\# FWHM line width (km/s)}
    \StringTok{"line\_area"}\NormalTok{: [}\FloatTok{250.0}\NormalTok{, }\FloatTok{125.0}\NormalTok{, }\FloatTok{175.0}\NormalTok{], }\CommentTok{\# line area (K km/s)}
    \StringTok{"velocity"}\NormalTok{: [}\OperatorTok{{-}}\FloatTok{35.0}\NormalTok{, }\FloatTok{10.0}\NormalTok{, }\FloatTok{55.0}\NormalTok{], }\CommentTok{\# velocity (km/s)}
    \CommentTok{\# normalized baseline coefficients}
    \StringTok{"baseline\_observation\_norm"}\NormalTok{: [}\OperatorTok{{-}}\FloatTok{0.5}\NormalTok{, }\OperatorTok{{-}}\FloatTok{2.0}\NormalTok{, }\FloatTok{3.0}\NormalTok{], }
\NormalTok{\})}

\CommentTok{\# Pack data structure with synthetic "observation"}
\NormalTok{observation }\OperatorTok{=}\NormalTok{ SpecData(}
\NormalTok{    velocity\_axis, sim\_brightness, noise,}
\NormalTok{    ylabel}\OperatorTok{=}\VerbatimStringTok{r"Brightness Temperature $T\_B$ (K)"}\NormalTok{,}
\NormalTok{    xlabel}\OperatorTok{=}\VerbatimStringTok{r"LSR Velocity $V\_\{\textbackslash{}rm LSR\}$ (km s$\^{}\{{-}1\})$"}\NormalTok{,}
\NormalTok{)}
\NormalTok{data }\OperatorTok{=}\NormalTok{ \{}\StringTok{"observation"}\NormalTok{: observation\}}

\CommentTok{\# Initialize the model with the synthetic observation}
\NormalTok{model }\OperatorTok{=}\NormalTok{ GaussModel(data, n\_clouds}\OperatorTok{=}\DecValTok{3}\NormalTok{, baseline\_degree}\OperatorTok{=}\DecValTok{2}\NormalTok{)}
\NormalTok{model.add\_priors()}
\NormalTok{model.add\_likelihood()}

\CommentTok{\# Draw posterior samples via MCMC}
\NormalTok{model.sample()}

\CommentTok{\# Solve labeling degeneracy}
\NormalTok{model.solve()}

\CommentTok{\# Draw posterior predictive samples}
\ImportTok{from}\NormalTok{ bayes\_spec.plots }\ImportTok{import}\NormalTok{ plot\_predictive}

\NormalTok{posterior }\OperatorTok{=}\NormalTok{ model.sample\_posterior\_predictive(thin}\OperatorTok{=}\DecValTok{100}\NormalTok{)}
\NormalTok{axes }\OperatorTok{=}\NormalTok{ plot\_predictive(model.data, posterior.posterior\_predictive)}
\NormalTok{axes.ravel()[}\DecValTok{0}\NormalTok{].figure.show()}

\CommentTok{\# visualize posterior distribution}
\ImportTok{from}\NormalTok{ bayes\_spec.plots }\ImportTok{import}\NormalTok{ plot\_pair}

\NormalTok{axes }\OperatorTok{=}\NormalTok{ plot\_pair(}
\NormalTok{    model.trace.solution\_0,}
\NormalTok{    model.cloud\_deterministics,}
\NormalTok{    labeller}\OperatorTok{=}\NormalTok{model.labeller,}
\NormalTok{)}
\NormalTok{axes.ravel()[}\DecValTok{0}\NormalTok{].figure.show()}
\end{Highlighting}
\end{Shaded}

\begin{figure}
\centering
\includegraphics{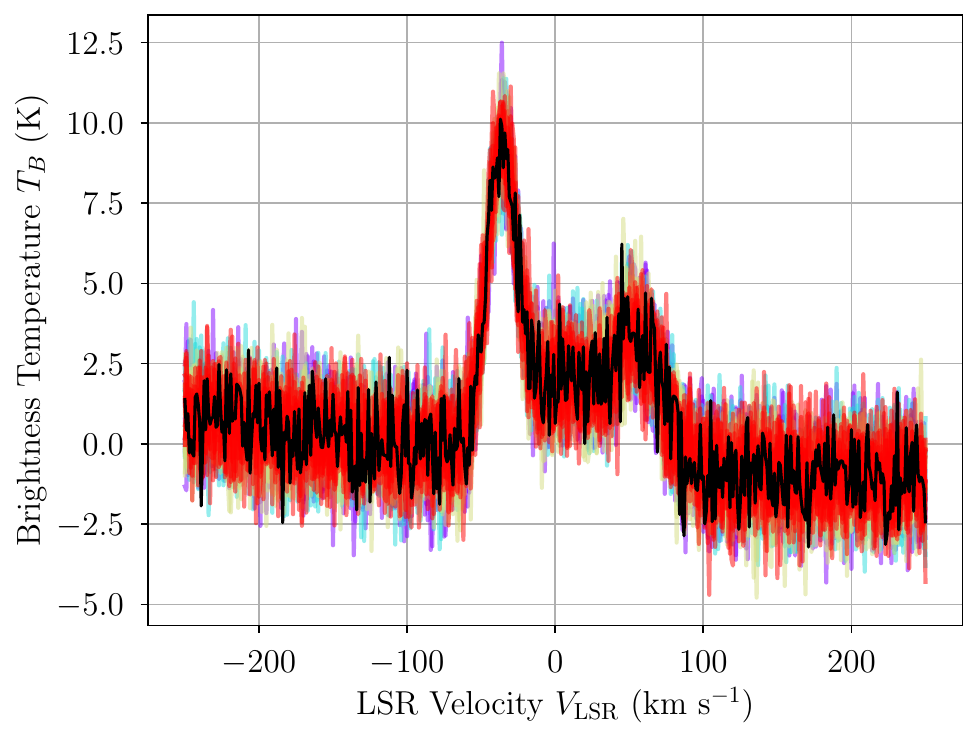}
\caption{Posterior predictive samples for a three-cloud
\texttt{GaussLine} model fit to a synthetic spectrum. The black line
represents the synthetic spectrum, and each colored line is one
posterior predictive sample.}
\end{figure}

\begin{figure}
\centering
\includegraphics{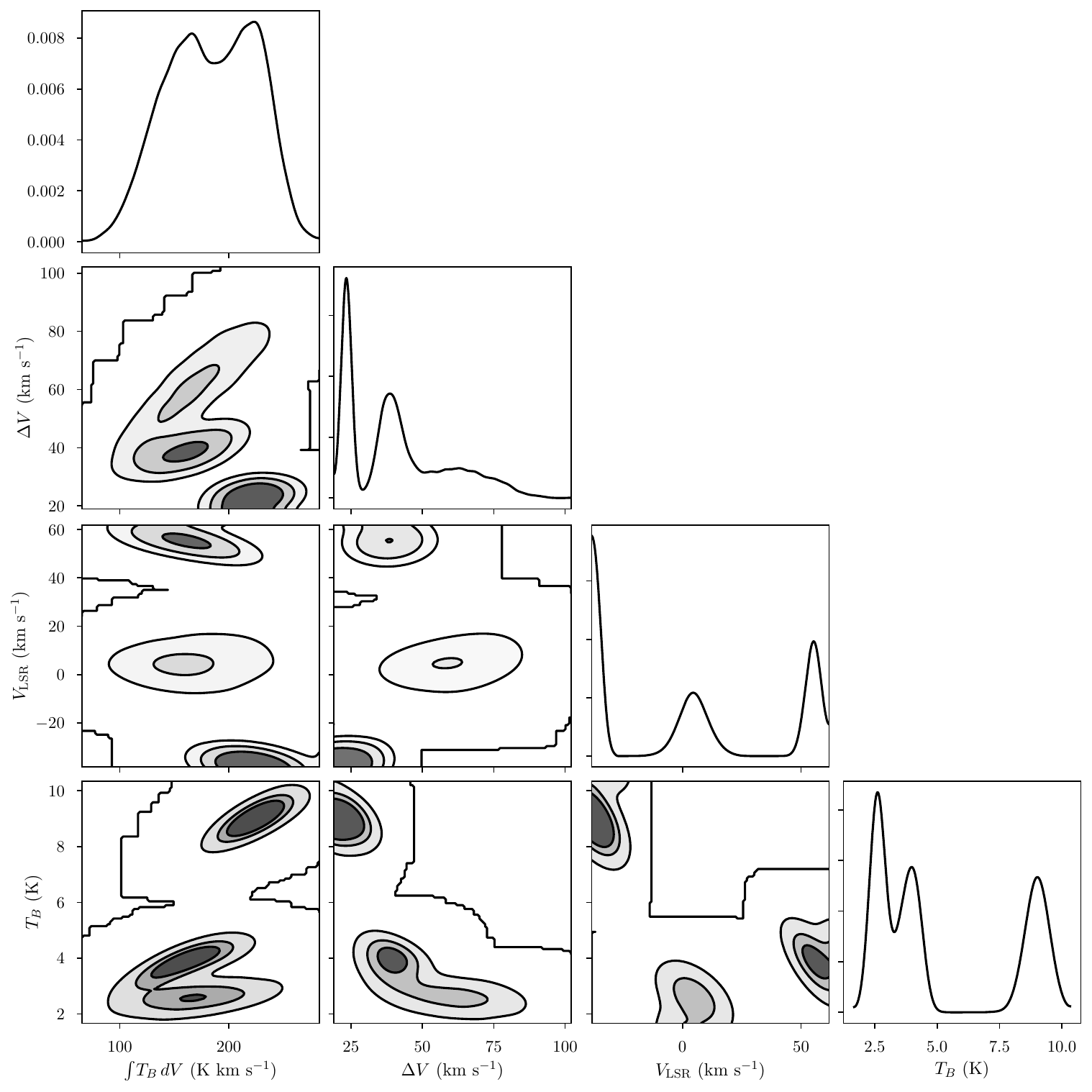}
\caption{Projections of the posterior distribution for a three-cloud
\texttt{GaussLine} model fit to a synthetic spectrum. The free model
parameters are the integrated line area, \(\int T_B dV\), the full-width
at half-maximum line width, \(\Delta V\), and the line-center velocity,
\(V_{\rm LSR}\). The line amplitude, \(T_B\), is a derived quantity. The
three posterior modes correspond to the three clouds in this model.}
\end{figure}

\section*{References}\label{references}
\addcontentsline{toc}{section}{References}

\phantomsection\label{refs}

\end{document}